\documentclass{aa}

\usepackage{graphicx}
\usepackage{txfonts}
\begin{document} 
\fancypagestyle{firstpage}
{
    \fancyhead{}
    \fancyfoot[L]{IPARCOS-UCM-25-012}
} 
   \title{The effect of mass and morphology on the mass assembly of galaxies}
   \titlerunning{Galaxy mass assembly and morphology}

   \author{A. Camps-Fariña
          \inst{1,2}\fnmsep\thanks{\email{arcamps@ucm.es}},
          R. M. M\'{e}rida\inst{3,4},
          P. S\'{a}nchez Blázquez\inst{1,2},
          \and
          S. F. S\'{a}nchez\inst{4}
          }
    \authorrunning{A. Camps-Fariña et al.}

   \institute{Departamento de F\'{i}sica de la Tierra y Astrof\'{i}sica, Universidad Complutense de Madrid, Pl. Ciencias, 1, Madrid, 28040, Madrid, Spain
   \and
   Instituto de Física de Partículas y del Cosmos, Universidad Complutense de Madrid, Pl. Ciencias, 1, Madrid, 28040, Madrid, Spain
   \and
   Centro de Astrobiolog\'{i}a,  (CAB), CSIC-INTA, Carretera de Ajalvir km 4, E-28850 Torrej\'{o}n de Ardoz, Madrid, Spain
   \and
   Departamento de F\'isica Te\'orica, Universidad Aut\'onoma de Madrid, E-28049, Cantoblanco (Madrid), Spain
   \and
   Instituto de Astronom\'{i}a, Universidad Nacional Aut\'{o}noma de M\'{e}xico, A.P. 106, Ensenada 22800, BC, M\'{e}xico
             }

 
  \abstract
{The pace at which galaxies grew into their current stellar masses and how this growth is regulated is still not fully understood, nor is the role that morphology plays in this process. We applied full spectral fitting techniques with \texttt{pyPipe3D} to the MaNGA sample to obtain its star formation and stellar mass histories and used these to investigate the mass assembly of galaxies by measuring how their specific star formation correlates to their stellar mass at different look-back times.

We find that the correlation between these two parameters was shallower in the past. Galaxies used to have similar mass doubling times and the current negative correlation between the specific star formation and M$_\star$ is primarily due to more massive galaxies 'dropping' off the main sequence earlier than less massive ones.
Additionally, selecting the galaxies into bins based on their present-day morphology shows a segregation in specific star formation rate (sSFR) that is maintained even at high look-back times, showing that the factors that determine which morphology a galaxy ends up in are in place at very early times. Similarly, selecting them based on their current star formation status shows that, on average, currently retired galaxies used to have slightly a higher sSFR before the drop-off, whereas galaxies that have continued to form stars until today had a lower sSFR initially.
We compare our results to a set of cosmic surveys, finding partial agreement in our results with several of them, though with significant offsets in redshift. Finally, we discuss how our results fit with certain theoretical models on galaxy evolution as well as cosmological simulations.}
   \keywords{galaxy evolution -- chemical abundances -- star formation
               }

   \maketitle
%

\section{Introduction}
It has been shown that the global star formation rate (SFR) has steadily decreased over cosmic time from z$\sim$2-3 \citep[][]{Madau2014}, so that galaxies today generally form fewer stars than in the past. However, this decline did not occur at the same rate for all galaxies. On average, the SFR in massive galaxies reached their highest value at a higher redshift and declined faster than in less massive galaxies, many of which are still forming stars \citep[e.g.][]{Speagle2014,Rodriguez-Puebla2017,Sanchez2019}. These differences lead to a number of correlations between the stellar mass (M$_\star$) and the mean age \citep[e.g.][]{Gallazzi2005,Blanton2009,Sanchez2018} and metallicity \citep[e.g.][]{Trager2000,Savaglio2005,Erb2006, Sanchez-Blazquez2006a,Sanchez-Blazquez2006b,Maiolino2008,Camps-Farina2021,Camps-Farina2022}. Moreover, the M$_\star$ function of galaxies (when observed at different redshifts) shows much smaller changes in the high M$_\star$ range than at the low end of the scale \citep[e.g.][]{Fontana2004,Bundy2005,Pozzetti2010}. These results are known as the downsizing phenomenon \citep{Cowie1996}, which generally suggests that more massive galaxies assemble their mass in shorter timescales. Despite all of the advances made in understanding these processes, we still do not know the details of the transition between star-forming and retired galaxies, what triggers that transformation, when it happens, and what regulates it.

To explore this issue, two different methods could be adopted: (i) study the SFR and M$_\star$ of different cosmological samples at a different redshift and compare them or (ii) derive the star formation histories (SFHs) of individual galaxies based on their current observed properties -- for example, spectral energy distribution (SED) fitting -- and infer their properties at different cosmological times.
The method we used to calculate the SFH for the considered sample of analysed galaxies is full spectral fitting: a technique that allows stellar populations to be dated within a galaxy \citep[e.g. see][]{Sanchez2020,Sanchez2021b}{}{}. Early studies used the central spectra of galaxies from the Sloan Digital Sky Survey (SDSS) \citep[e.g.][]{York2000}, but in recent years the advent of integral field unit (IFU) surveys such as CALIFA \citep{Sanchez2012}, SAMI \citep{Bryant2015}, and MaNGA \citep{Bundy2015} has allowed for more robust determinations using the integrated spectrum of the galaxies \citep[e.g.][]{Sanchez-Blazquez2011,CidFernandes2013,Sanchez-Blazquez2014,Ibarra-Medel2016, Goddard2017}. \cite{Sanchez2019} studied the star formation main sequence \citep[SFMS,][]{Brinchmann2004,Noeske2007,Elbaz2011} over cosmic time and reproduced the results of \cite{Madau2014} using stellar population synthesis techniques.

The way that galaxies transition from being in the main sequence of star formation until they become retired galaxies is still not well understood. There are many different mechanisms that have been proposed and appear to correlate well with the quenching of the galaxies, such as feedback from active galactic nuclei (AGN) or galactic outflows \citep[e.g.][]{Dekel2006,Sturm2011,Gaspari2012,Cicone2014,Sanchez2018, Dekel2019,Trussler2020}, morphological quenching ,\citep[e.g.]{Martig2009, Schawinski2014, Smethurst2015}, or environment \citep[e.g.][]{Wetzel2012, Boselli2014}, as the main actors. It is thus very likely that quenching effects act on global scales \citep[e.g.][]{Belfiore2017,Belfiore2018, Trussler2020} and involve a loss of gas supply \citep[e.g.][]{Boselli2006,Dekel2006, Schaye2010,Dekel2019, Trussler2020}.

As indicated before, global trends in SFHs can also be studied using cosmic surveys in which galaxies are observed at various redshifts. While these studies lose the advantage of measuring the evolution of the SFR on the same sample of galaxies, they also avoid the uncertainties and degeneracies present in spectral fitting techniques \citep[][]{Worthey1999,Kuntschner2001,Sanchez-Blazquez2011,Walcher2011,Conroy2013, CidFernandes2014}.
Deep observations of cosmological fields on the sky, in particular the \textit{Hubble} Deep Field and the Great Observatories Origins Deep Survey fields \citep[GOODS;][]{Giavalisco2004, Dickinson2003} with the \textit{Hubble} Space Telescope (HST), have enabled a large number of surveys to measure the SFR and M$_\star$ of galaxies over a wide range of redshifts \citep[e.g.][]{Salim2007,Noeske2007,Daddi2007,Perez-Gonzalez2008b,Stark2009, Karim2011,Grogin2011,Koekemoer2011,Schreiber2015,Ilbert2015,Salmon2015,Tomczak2016,Santini2017}, which are currently being dramatically expanded with the introduction of the \textit{James Webb} Space Telescope \citep[][]{Gardner2006}.

These surveys generally agree well with the general picture of the evolution of the global SFR and SFMS described above \citep[][]{Speagle2014}, but they have significant uncertainties in the absolute values of the SFR and M$_\star$. As cosmic surveys employ several methods and they measure galaxies, by design, at vastly different distances, it is also difficult to distinguish features of the SFR- or sSFR-M$_\star$ relations beyond the slope and zero-point and how these evolve through time. Morphology is also a complicated parameter to include, both due to observational constraints in spatial resolution given the distances involved, and the fact that it is not constant through time \citep[][]{Conselice2014}. This is precisely where the full spectral fitting results shown in this paper can contribute to the understanding of how galaxies grow their mass, as the fact that our sample is composed of the same galaxies at all cosmic times allows us to probe qualitative trends intrinsic to the galaxies regarding how the sSFR-M$_\star$ relation has evolved in cosmic time.

Specifically, we investigate how star formation is regulated in galaxies by examining which parameters determine the shape and evolution of the relation between specific star formation and stellar mass (sSFR-M$_\star$) with cosmic time. Using the MaNGA galaxy sample and applying stellar population synthesis techniques, we could track the evolution of the same galaxies over cosmic time and partition them into current morphology bins that are consistent over cosmic time because they contain the same galaxies that end up in a given morphology bin at all times. We also checked whether galaxies that are currently forming stars show a different evolution than galaxies that are retired. The results are compared with several cosmic surveys adapted to the format of our results.

The text is organised as follows: In Sec. \ref{sec:data} we describe the two types of data used, IFU observations, and a spectrophotometric survey; in Sec. \ref{sec:sample} we describe the sample employed and how it was selected; in Sec. \ref{sec:analysis} we describe the reduction and analysis procedures used to obtain the relevant quantities; and in Sec. \ref{sec:results} we show the resulting relations between the sSFR(t), M$_\star$(t), current SFR, and current morphology. In Sec. \ref{sec:discussion} we discuss the implications of the results and their validity before presenting the conclusions in Sec. \ref{sec:conclusions}.

\section{Data}\label{sec:data}
Mapping Nearby Galaxies at APO (MaNGA) \citep{Bundy2015} is a project in the fourth iteration of SDSS \citep[][]{York2000} to produce maps of the ionised gas and stellar properties for a large number of galaxies in the Local Universe ($\langle \mathrm{z} \rangle \sim 0.03$) as a statistically significant sample to study how galaxies have evolved over their lifetime. Observations were made using a set of fibre bundles fed into the Baryon Oscillation Spectroscopic Survey (BOSS) spectrographs \citep{Smee2013} to map the spectral emission of the galaxies at each spatial position, aiming to cover most of the area of each galaxy. The coverage primary sample (66\%) galaxies includes at least 1.5 effective radii (R$_\mathrm{e}$) and for the secondary sample (33\%) it includes 2.5 R$_\mathrm{e}$. The spectrographs are fed by optical fibres which are bundled to produce integral field units (IFUs) that provide spatially resolved spectral information. The IFUs consist of a primary bundle that can range from 19 (with a diameter of 12\arcsec) to 127 fibres (with a diameter of 32\arcsec), depending on the object, and 12 sets of seven fibre bundles for flux calibration and 92 fibres for sky subtraction \citep{Drory2015, Yan2016}. The fibre bundles are inserted into precision perforated plates prepared for each night's observing targets and then mounted on the Apache Point Observatory 2.5-m Sloan Telescope at Apache Point Observatory \citep{Gunn2006}.

The raw data from the telescope are processed with the MaNGA data reduction pipeline \citep[DRP,][]{Law2016} into the project's core products (the IFU data cubes that we use here).
Each data cube has a spectral range between 3600 \AA{} and 10,300 \AA{} at a resolution of R $\sim$ 2000, with an effective spatial resolution after reduction of about 2.5\arcsec/FWHM.

\section{Sample}\label{sec:sample}
The MaNGA sample consists of $\sim$ 10,000 galaxies selected in the Local Universe with the primary criteria being that (i) they are a representative sample of the local galaxy population and (ii) they are well covered by the field of view (FOV) of the instrument. The latter is necessary to obtain 2D maps on several properties such as M$_\star$ density, mean stellar age, SFR, metallicity, etc.

We apply some selection criteria to the parent sample to obtain a sub-sample suitable for our study, removing galaxies with an inclination above 70º and those in which an AGN has been detected \citep[from][]{Lacerda2020}. Measurements in highly inclined galaxies are heavily affected by dust lanes and it is generally harder to obtain quality SFHs, while wide emission lines from AGN can also affect the analysis whilst being a relatively small number of galaxies. The refined sample consists of 9087 galaxies and is the same as that used in \cite{Camps-Farina2022}.

We use the \cite{Vazquez-Mata2022} morphological classification (v2.0.1), which is available as a Value Added Catalog\footnote{\url{https://www.sdss4.org/dr15/data_access/value-added-catalogs/?vac_id=manga-visual-morphologies-from-sdss-and-desi-images}}. The classification procedure uses r-band SDSS and Dark Energy Spectroscopic Instrument \citep[DESI,][]{Dey2019} images which are processed with a sharpening filter and have surface brightness models subtracted to detect sub-structures such as bars.

The main morphology bins used in this article can be reproduced by selecting the following morphology labels from the Value-Added Catalog:
\begin{itemize}
    \item \textbf{E-S0}: E, E(dSph), S0, S0(dwarf), S0a, SAB0, SAB0a, SB0, SB0a
    \item \textbf{Sa-Sb}: SABa, SABab, SBa, SBab, Sa, Sab, SABb, SABbc, SBb, SBbc, Sb, Sbc
    \item \textbf{Sc-Irr}: SABc, SABcd, SBc, SBcd, Sc, Scd,  Irr, IrrAB, IrrB, SABd, SABdm, SBd, SBdm, Sd, Sdm, dIrr, dSph, dwarf
\end{itemize}

In other words, we segregate galaxies into: (i) early-type ones (bulge dominated), (ii) early spirals (galaxies with both bulges and discs), and (iii) late spirals (galaxies without a significant bulge, mostly disc dominated).

\section{Analysis}\label{sec:analysis}
The composite optical-near-infrared stellar spectrum of a given region in a galaxy varies as a function of the SFH, the metallicity of the stars, and the present-day dust extinction. The process of fitting a library of stellar population spectra to observations to determine the composition of the underlying stellar population allows us to study the past of a galaxy using features left behind in the current stellar populations.

\subsection{\texttt{pyPipe3D}}\label{sec:pipe3d}

\texttt{pyPipe3D} \citep[][]{Lacerda2022} is an analysis pipeline used to obtain the ionised gas and stellar population parameters of IFU observations such as emission line fluxes, velocities and widths and stellar population ages and metallicities, among others. It also provides us with the relevant 2D maps of these parameters. The stellar spectra are fitted using templates from the library of simple stellar populations (SSPs) specified by the user, returning the light fractions of the populations for each given spectrum.
It is an updated version of the Pipe3D analysis pipeline \citep{Sanchez2006,Sanchez2016a,Sanchez2016b} rewritten in \texttt{Python}, with added modularity and ease of use, and a significant increase in computational speed. It separates the emission and absorption components and fits them separately to measure and generate spatially resolved maps of the relevant physical quantities of the galaxy.
Internal dust attenuation is determined using a two-step procedure as extensively described in \cite{Lacerda2022} and \cite{Sanchez2016a}. The procedure performs an estimation of the dust attenuation by doing a brute force exploration in a fixed range of values, fitting the stellar population using a limited subset of the stellar template that limits the possible degeneracies between properties. A \cite{Cardelli1989} extinction law is applied.

\subsection{Obtaining the star formation relations}\label{sec:ssfr_t50}
The M$_\star$ of a galaxy can be determined from the light fractions provided by \texttt{pyPipe3D} as follows:
\begin{equation}
M_\star = \sum_{i}^{N_{age}}\sum_{j}^{N_Z} \, w_{i,j} \, \Upsilon_{i,j} \, L_{V,tot}
\end{equation}

where $w_{i,j}$ and $\Upsilon_{i,j}$ are the light fraction and the mass-to-light ratio of the SSP template with a given age and metallicity, respectively. We can obtain the mass assembly history (MAH), which is the mass of all the stars present at a given age of the Universe ($t_U = 13.7 - t$) as:

\begin{equation}
M_\star (t) = \sum_{i > t}^{N_{age}}\sum_{j}^{N_Z} \, \left (1+F_{loss}(i - t)\right) \, w_{i,j} \, \Upsilon_{i,j} \, L_{V,tot}
\end{equation}

where $F_{loss}(i - t)$ is the fraction of M$_\star$ returned to the ISM by a SSP of a given age and metallicity in the time interval between $t$ and now ($t=0$).
The sSFR is defined as the SFR divided by the M$_\star$ so we can trivially obtain the sSFR history with the SFH and the M$_\star (t)$.

Due to the nature of the sample selection performed for MaNGA, which prioritised a good coverage of the FOV, there is a significant redshift range for these objects (z$\sim$0.0002-0.15) which correlates with M$_\star$. This results in more massive galaxies  being observed at slightly earlier epochs compared to less massive ones.

We can consider the SFR for each look-back time (LBT) as equivalent to a measurement of a single object at a given redshift in a cosmic survey, similar to the methodology used in \cite{Sanchez2019}.
We thus select several discrete LBT values and measure the SFR and M$_\star$ of each galaxy at each time point. We do this by interpolating the SFH and M$_\star(t)$ curves, since the SFH of each galaxy is shifted by the value of its redshift and therefore, the ages of the templates correspond to slightly different LBT values. The lowest LBT value for which we measure the sSFR-M$_\star$ relation is 1 Gyr. This is due to the aforementioned redshift range in the sample, which makes it so that for very low LBT values many galaxies are not observed as the light travel time is longer than 1 Gyr for the distance they are located at. In other words,  for some galaxies the light we currently measure was emitted longer than 1 Gyr ago due to their distance.
At 1 Gyr 87\% of the sample is included, which we considered a good compromise for picking the lowest LBT value.

\section{Results} \label{sec:results}
\begin{figure}
        \includegraphics[width=\columnwidth]{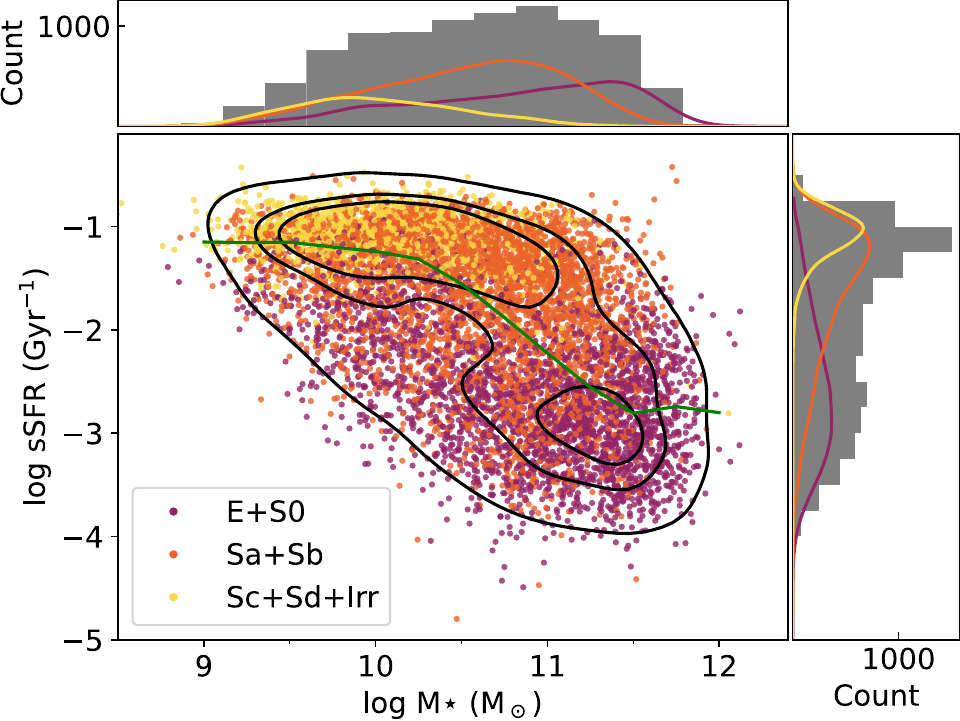}
    \caption{Currently observed sSFR of MaNGA galaxies compared to their M$_\star$ (bottom left). The black contours enclose 95\%, 65\% and 35\% of the sample respectively. The green line is the median sSFR within 0.25 dex wide M$_\star$ bins. In the top and right panels, the distributions in M$_\star$ and sSFR, respectively, are shown as histograms, with the three morphological bins shown in lines of the corresponding colour.}
    \label{fig:sfms_current}
\end{figure}
In Fig. \ref{fig:sfms_current} we show the sSFR-M$_\star$ relation for the galaxies in our sample. The SFR is measured by averaging it over the last 30 Myr which is a typical value for the lifetime of an HII region and this is also the timescale at which the SFR measured using H$\alpha$ and full spectral fitting coincide \citep{Asari2007,GonzalezDelgado2016}. Similarly to the SFMS, there is a bi-modal distribution with a fairly flat upper distribution of star-forming galaxies and a lower distribution of retired galaxies, which are connected by the Green Valley.

\begin{figure*}
        \includegraphics[width=\linewidth]{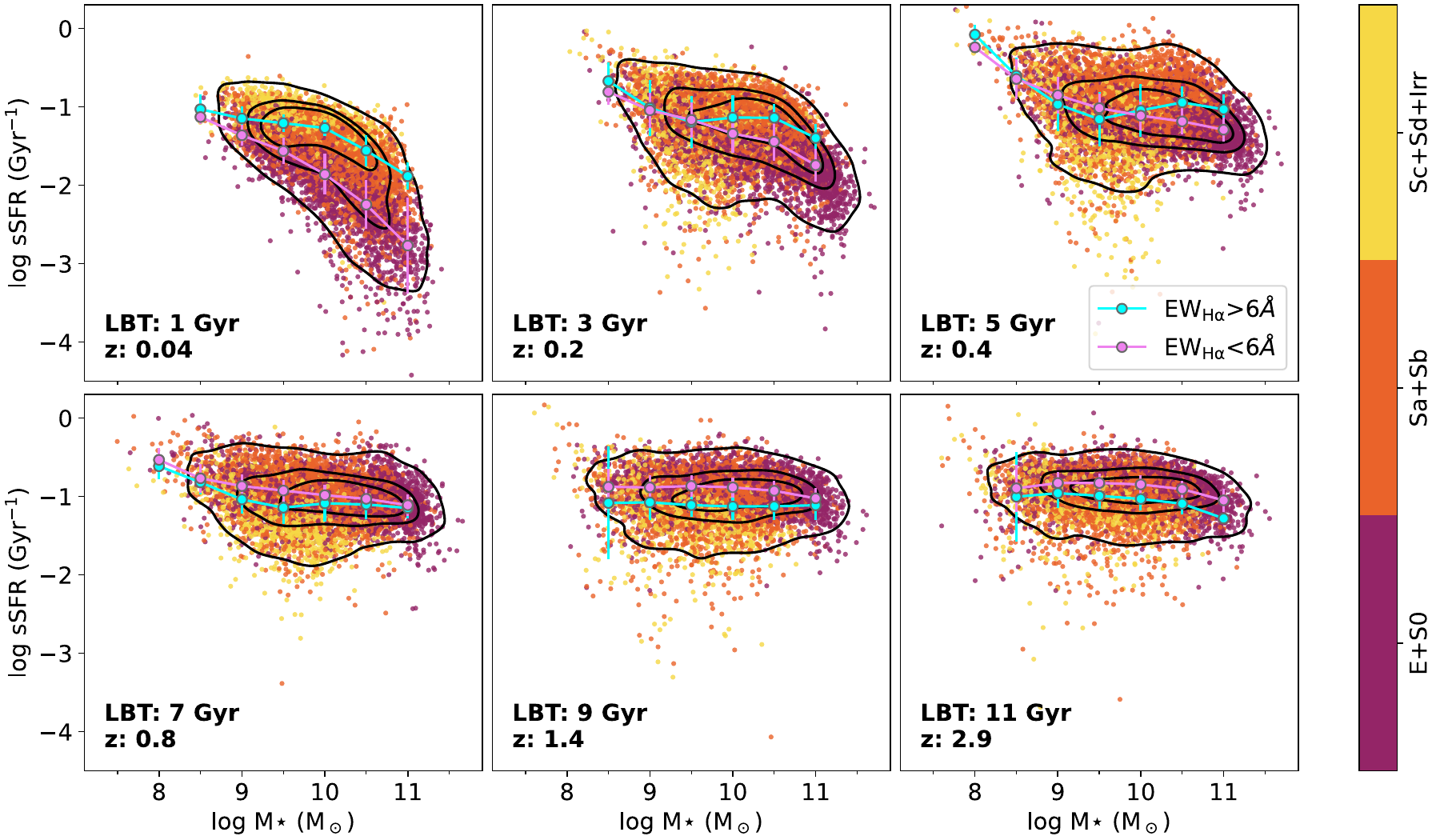}
    \caption{Evolution of the sSFR over cosmic time for the galaxies in the MaNGA sample. In each panel the M$_\star$ and sSFR of the galaxies measured at different LBT are shown. The colour of the data points corresponds to the currently observed morphology of the galaxies. The black contours enclose 95\%, 65\% and 35\% of the sample, respectively. The cyan and violet dots and lines show the median sSFR within 0.5 dex wide M$_\star$ bins, selecting galaxies above and below 6 \AA{} in EW$_\mathrm{H\alpha}$, respectively. The error bars for the dots show the 25th and 75th percentile of the distribution within each M$_\star$ bin.}
    \label{fig:ssfms_dist}
\end{figure*}

In Fig. \ref{fig:ssfms_dist} we show the sSFR-M$_\star$ distribution measured at different LBT by evaluating the SFH and MAH at these LBT values. At 1 Gyr, the relation is very similar to that shown in Fig. \ref{fig:sfms_current}, with a general negative trend between the two parameters which steepens at about log M$_\star$/M$_\odot$ = 10.5. The retired galaxies no longer have a clearly defined separate distribution, likely the effect of measuring at a specific LBT rather than the last 30 Myr, meaning that for a portion of these galaxies the SFH is measured at the time when they are still becoming quenched, widening the distribution.

We show averaged sSFR values of the distribution within M$_\star$ and current star formation bins as large dots on Fig. \ref{fig:ssfms_dist}. The M$_\star$ bins are selected at each LBT but the star-forming (SFG) and retired (RG) galaxy bins are selected based on whether the current equivalent width in H$\alpha$ (EW$_\mathrm{H\alpha}$) of the galaxies is higher or lower, respectively, than 6\AA{} \citep{Cano-Diaz2016}. Each bin that is represented has at least ten objects in it.
SFG and RG are still segregated at 1 Gyr and they also have different sSFR-M$_\star$ slopes such that at higher M$_\star$ RG have much lower sSFR compared to SFG. The bend at log M$_\star$/M$_\odot$ = 10.5 is mainly observed in SFG rather than RG. The sSFR-M$_\star$ relation for both star formation bins appears to converge at low masses such that they have similar sSFR at 1 Gyr in LBT, despite having different EW$_\mathrm{H\alpha}$ values. The most likely explanation for this is that low M$_\star$ galaxies have their star formation in short, intermittent bursts \citep[e.g.][]{Lee2009,Dominguez2015,Atek2022}. As such, the emission line based separation using EW$_\mathrm{H\alpha}$ would not correlate with a significant separation at 1 Gyr. Galaxies which are currently star-forming are not as statistically likely to be so once we look at the recent past if they tend to have short bursts of star formation.

The relation between the sSFR and M$_\star$ becomes flatter at higher redshifts, indicating that the SFR used to be roughly proportional to the M$_\star$ at earlier times, which creates a flat relation between M$_\star$ and the sSFR. A flat correlation means that galaxies double their M$_\star$ on similar timescales and the currently observed negative correlation means that more massive galaxies take longer to double their M$_\star$.
As such, we propose that the phenomenon of downsizing is not best described as more massive galaxies having a quicker evolution than less massive galaxies. Instead, the self-regulation of the SFR means that galaxies grow at a similar pace (similar M$_\star$ doubling times) intrinsically but once galaxy-wide quenching mechanisms become efficient their growth is stopped, a phenomenon that happens earlier in more massive galaxies. Therefore, it is not that the evolution of the latter is intrinsically quicker (defined by M$_\star$-doubling times) but that their steady-state of star formation was shut down at an earlier time than less massive galaxies. As such, downsizing is defined by the age at which quenching mechanisms become efficient.

The difference between star-forming and retired galaxies (in the Local Universe) quickly shrinks with increasing LBT, showing once more that these galaxies used to form stars at a similar pace before some of them shut off further star formation. The timing for currently retired galaxies to become quenched depends on M$_\star$, with less massive galaxies reaching similar sSFR values at 3 Gyr in LBT (z$\sim$0.2) but more massive ones only do so at 5 Gyr (z$\sim$0.4) and earlier. Interestingly, there is a small but consistent gap between RG and SFG at the highest LBT of about 0.11-0.23 dex, showing that currently retired galaxies had slightly higher sSFR at the beginning. This implies that galaxies that are retired in the present day are statistically more likely to have been highly efficient at forming stars in the early universe. As such, there is a degree of intrinsic downsizing such that some galaxies did have a quicker evolution, but effect does not appear to be strongly tied to M$_\star$, unlike the global quenching described above.

At 3-7 Gyr in LBT (z$\sim$0.2-0.8) the lowest M$_\star$ galaxies show a deviation from the general trend such that the relation between sSFR and M$_\star$ significantly steepens, especially at 5 Gyr (z$\sim$0.4). The number of galaxies involved is small relative to the full sample except at 5 Gyr, where the change in slope occurs at around log M$_\star/$M$_\odot \sim 9-9.5$. Were this a real feature, it would imply that galaxies with log M$_\star$/M$_\odot \lesssim 8.5-9$ have the opposite trend to that observed for galaxies above this M$_\star$, that is, that for earlier LBT the relation between sSFR-M$_\star$ becomes steeper instead of shallower.
\subsection{Morphology} \label{sec:ssfr_morph}
In this section we explore if we can distinguish galaxies with different morphologies according to their SFH and how far into the past this segregation is maintained.
The morphology of galaxies in the local universe is closely related to their SFH \citep[e.g.][]{ Camps-Farina2021, Camps-Farina2022}. Thus, to some extent, we should expect to find additional correlations if we divide the sample into morphology bins before computing the evolution of the sSFR-M$_\star$ relationship. At the very least, we should find a more pronounced decline in sSFR over cosmic time for early-type galaxies, whose retired fraction is much higher than for late-type galaxies, which tend to have ongoing star formation.

A crucial difference for the comparison between morphology bins is that, unlike M$_\star$ bins, we are not able to distinguish the morphology of galaxies at different cosmic times. For this reason, these morphology bins must be interpreted as representing galaxies that end up in a particular morphological bin. In addition, the current morphology classification is unlikely to be representative of the galaxy population at all cosmic times \citep[e.g.][]{Conselice2014}.

\begin{figure*}
        \includegraphics[width=\linewidth]{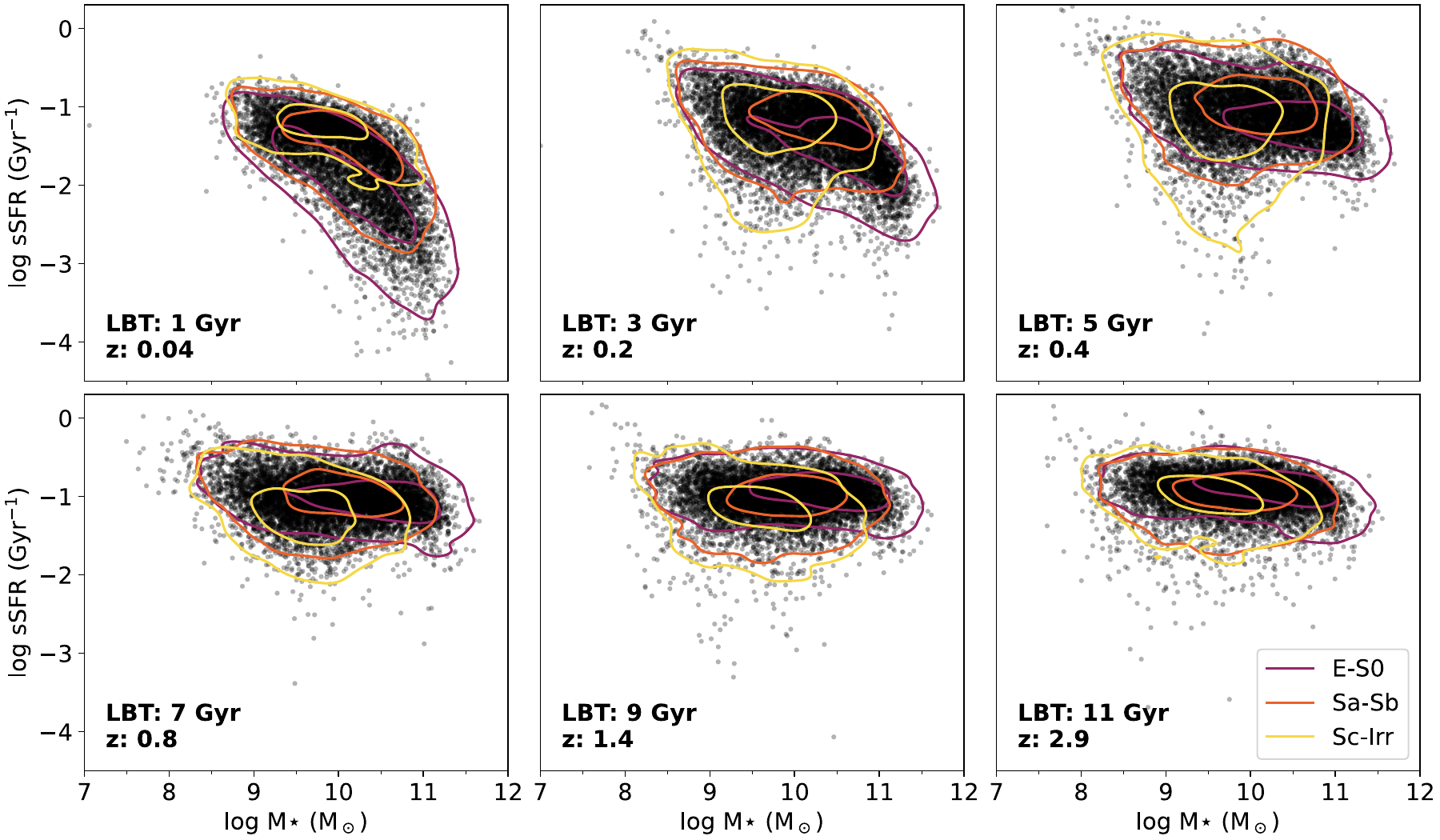}
    \caption{Evolution of the sSFR over cosmic time for the galaxies in the MaNGA sample. In each panel the M$_\star$ and sSFR of the galaxies measured at different LBT are shown. There are three sets of contours whose colours correspond to the current morphology of the galaxies and whose levels enclose 95\% and 50\% of each sub-sample.}
    \label{fig:ssfms_morph}
\end{figure*}

In Fig. \ref{fig:ssfms_morph}, we show the same data points as in Fig. \ref{fig:ssfms_dist} but we overlay the contours of galaxies divided in three morphological bins: early type (E and S0), intermediate type (Sa and Sb), and late type (Sc, Sd, and Irr). 
At 1 Gyr in LBT the morphological types have clear differences in both the extent of the M$_\star$ range that they cover, their average sSFR and the slope of the correlation between M$_\star$ and sSFR. Early-type galaxies are widely distributed in M$_\star$ and have lower sSFR in general as well as having a steep negative correlation such that more massive galaxies have much lower sSFR than less massive ones. On the other hand, the M$_\star$ distribution of late-type galaxies is narrower and centred at lower values, and the correlation becomes flatter. Late-type galaxies show a fairly flat distribution while intermediate ones have a clear bend in the relation such that the slope increases above log M$_\star$/M$_\odot$ $\sim$ 10-10.5. The difference in slope between morphological bins results in a convergence of sorts at lower M$_\star$ such that at log M$_\star$/M$_\odot$ $\sim$ 9-10 the differences in sSFR due to morphology are relatively small.

The differences between morphological types at 1 Gyr are similar to those between SFG and RG seen in Fig. \ref{fig:ssfms_dist} as would be expected due to the correlation between star formation status and morphology. The relation between sSFR and M$_\star$ becomes shallower at earlier LBTs for all morphology bins, most clearly for E+S0 and Sa+Sb but even Sc-Irr (whose distribution is fairly flat at all times) can be seen to have a shallower relation at 11 Gyr compared to 1 Gyr. The main change in the latter morphological bin is that its distribution greatly widens between 3-7 Gyr, perhaps indicating a period of increased burstiness. Early-type galaxies maintain a shallow relation between sSFR and M$_\star$ for LBT$\sim$11 to 7 Gyr, but at 7 Gyr the centre of the distribution has already started to drop and by 5 Gyr it is clearly located at lower sSFR compared to the other morphological bins. This indicates that their star formation slows down at around 7 Gyr compared to galaxies in the other morphological bins before the current negative correlation between sSFR and M$_\star$ was established.

At high LBT we see that the galaxies in each of the current morphology bins have similar distributions, but they are distinctly segregated in M$_\star$, even more so than at 1 Gyr and especially at the lower M$_\star$ end of the distributions. Such a clear difference between the galaxies at such an early time despite the bins corresponding to only current properties of the galaxies shows that current morphology is determined by the entire history of a galaxy's evolution starting at the very beginning of its lifetime. It also suggests that currently retired galaxies are not, on average, simply star-forming galaxies with a recent and rapid quenching but that they tend to have different SFHs even at early times.

There is also a small (0.1 dex) separation between the average sSFR between early- and late-type galaxies, with early-type galaxies having slightly higher sSFR. We expect this segregation to be related to the one found at this same LBT between RG and SFG in Fig. \ref{fig:ssfms_dist}; however, the segregation is significantly smaller between the morphology bins despite us measuring the distance between the two extremes of three bins instead of the difference between two bins. As such, we can conclude that the segregation in sSFR at early times is more likely to be related to whether galaxies continue to have star formation in the present than their morphology.

Of course, the opposite interpretation is also valid: More massive galaxies which have high star formation efficiency (relative to M$_\star$) at early times are less likely to end up in later type morphologies compared to galaxies of similar M$_\star$ but with lower early sSFR. Either way, a clear link between M$_\star$ at early times and present-day morphology is established. This result points towards a possible link between large scale mechanisms that make a galaxy grow quicker than average in the early universe and the eventual morphology of the galaxies. One possible explanation would be that galaxies which are formed in deeper dark matter potential wells accrete more gas at early times producing an early growth, but it also makes them more likely to attract other galaxies to interact with, eventually favouring an early-type morphology as a result of the interactions. The existence of long-term differentiation of SFHs between galaxies has been proposed by \cite{Caplar2019,Matthee2019} from results in simulations.

The fact that the distributions are segregated more at low LBT than at high ones suggests a 'convergence' of sorts in the MAHs of galaxies of different morphology. In other words, galaxies with different morphology which have different M$_\star$ at very early times end up, on average, with M$_\star$ values which are more similar in later times. This implies that late-type galaxies have formed a higher percentage of their M$_\star$ in the 1-11 Gyr interval in LBT compared to early-type ones, which could be an effect of gas accretion. Cosmic gas is expected to (i) be preferentially accreted at the outskirts of galaxies \citep{SanchezAlmeida2014,SanchezAlmeida2017} and (ii) to form discs, making so that galaxies with lower initial M$_\star$ in our sample are statistically more likely to develop discs and therefore they are more likely to be late type currently. \cite{Camps-Farina2023} measures the gas accretion history of galaxies and finds a correlation such that late-type galaxies have higher gas accretion rates in their recent history, in line with this interpretation.

\subsection{Statistical significance of the populations}

The previous figures show how the different populations of galaxies in terms of M$_\star$, morphology and present-day star formation have different evolutionary paths well into their past. The differences appear to be fairly consistent between ages for all the galaxy populations, but it is still important to use proper statistical metrics to determine how significant their differences are and how far into the past do these populations remain distinct.

For this purpose, we have performed a Kolmogorov-Smirnov (KS) two-sample test for each relevant pair of distributions to see how likely it is that the two populations have the same origin. A p-value lower than 0.05 tells us that it is extremely unlikely that the two samples are the same.
The relevant pairs are: one for the difference between current SFG and RG, and three pairs for all the combinations between early-, intermediate-, and late-type galaxies. The distribution in log M$_\star$/M$_\odot$ within each population is different and therefore comparing the entire population could lead us to overestimate how different their distributions in sSFR are. As an example, consider a single sample of galaxies which have a correlation between log M$_\star$/M$_\odot$ and sSFR and no secondary correlations. If we divide the sample into a high-M$_\star$ one and a low-M$_\star$ one their average sSFR will be different due to the correlation with M$_\star$ and we would obtain a false result of two different samples.

\begin{figure*}
        \includegraphics[width=\linewidth]{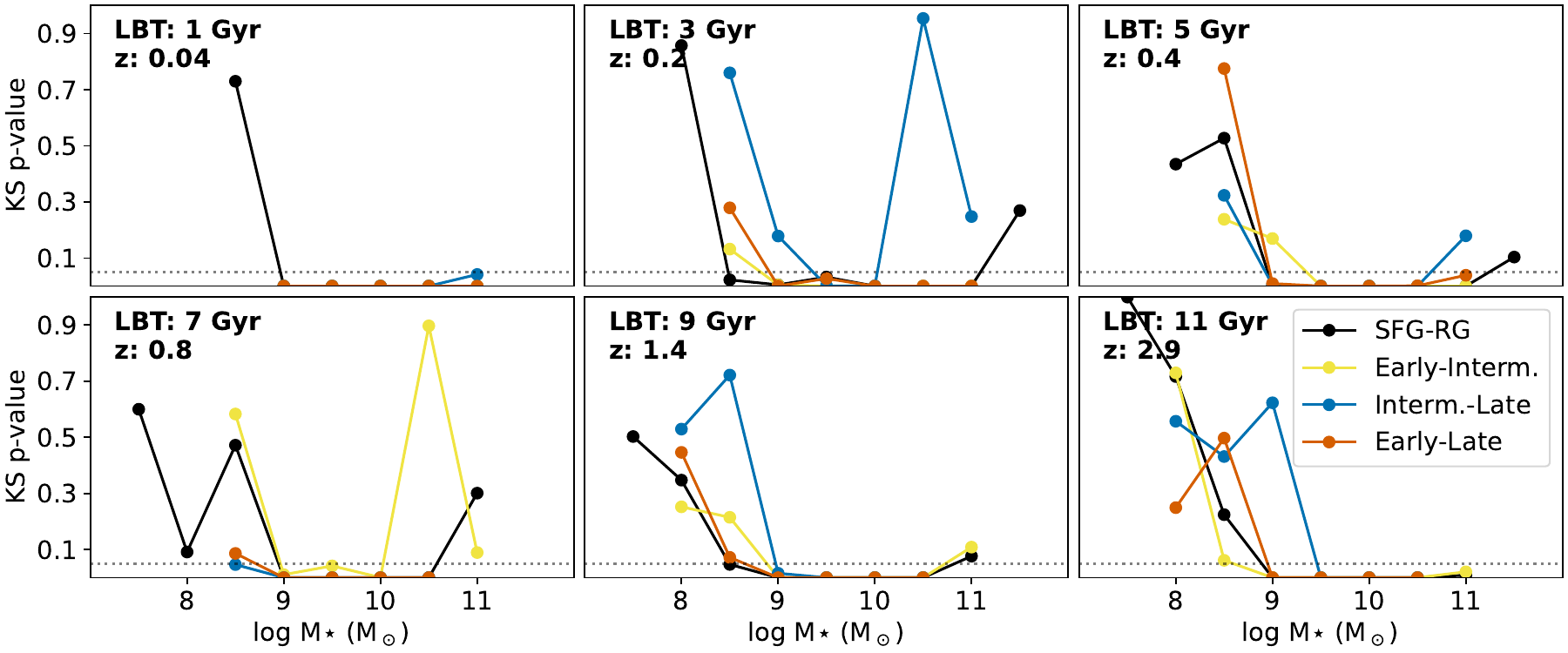}
    \caption{Two-sample Kolmogorov-Smirnov test for each pair of galaxy populations considered in the article. In each panel we show the p-value for each pair of populations performed individually at each of the 0.5 dex wide M$_\star$ bins. The dotted line shows the limit of p-value$\,= 0.05$, below which the samples are considered to be statistically significantly different. The colours identify the pairs of populations used.}
    \label{fig:KS_test}
\end{figure*}

To avoid this, we have performed the KS test within the 0.5 dex wide M$_\star$ bins employed in the averaged SFG and RG populations in Fig. \ref{fig:ssfms_dist}. This removes any dependence on log M$_\star$/M$_\odot$ as now we measure the p-value for the distributions of sSFR for each population within a given M$_\star$ bin. The resulting p-value vs log M$_\star$/M$_\odot$ inform us of which M$_\star$ ranges show significant differences in the distribution of sSFR values of the populations considered.
The resulting p-values show that for the M$_\star$-range where we have good statistical sampling, log M$_\star$/M$_\odot \sim 9-11$ the samples considered are clearly distinct, with values much lower than the 0.05 threshold. In fact, even at 11 Gyr in LBT the median p-value between log M$_\star$/M$_\odot = 9-11$ for all pairs of populations is $1.6 \cdot 10^{-15}$ indicating that, even at the highest age we consider, these populations remain very significantly distinct.

 There is also a possible interesting result showing that at log M$_\star$/M$_\odot$ = 10.5 there are two LBT values with an unusually high similarity in one of the morphology pairs. This is the M$_\star$ where the Green Valley in the SFMS appears and is considered a watershed mass for galaxies to start to quench \citep[e.g.][]{Dekel2006,Dekel2019}. The fact that we observe first an isolated very high similarity between the early and intermediate and then another one between intermediate and late (as seen in the present) could indicate the moment in which the currently early and the currently intermediate samples started to quench. Comparing this to Fig. \ref{fig:ssfms_morph} we do see that at these times and M$_\star$ the relevant pair of distributions appear to have similar values as the sSFR starts to drop for earlier-type morphology. This is not seen in the SFG-RG distributions and the morphology bins become distinct again in the next LBT, pointing to a transitional period with a $\lesssim 2$ Gyr timescale.
 
\subsection{Comparison with cosmological surveys}
In this section we compare our results with those of other studies which measure the evolution of either the SFR-M$_\star$ or the sSFR-M$_\star$ relations by measuring these parameters at different redshift values based of different galaxy surveys. Given the drastically different methodology employed by these studies and our own, it is very useful to compare them to see if our results agree with theirs.
\begin{figure*}
        \includegraphics[height=0.93\textheight]{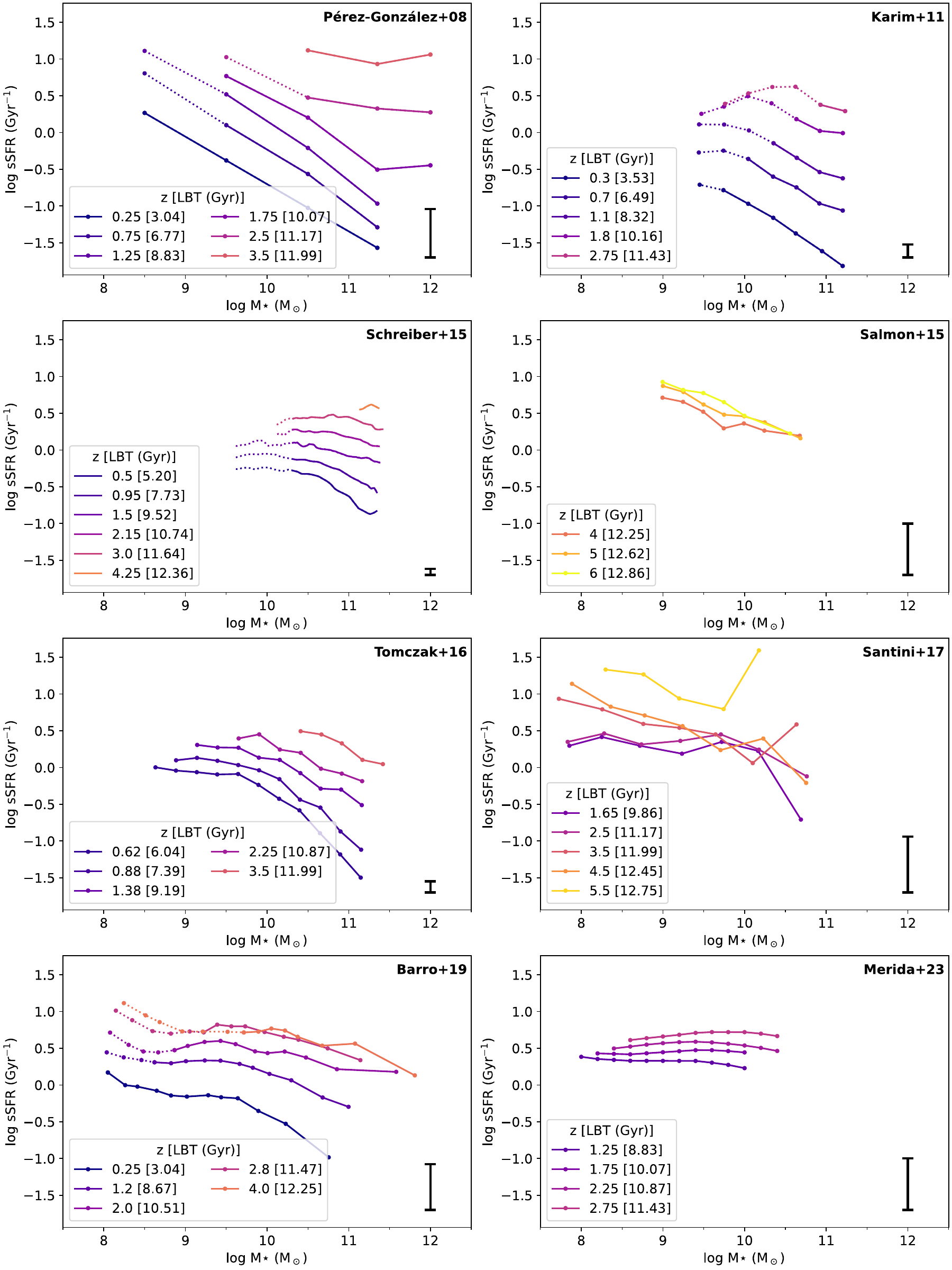} \caption{sSFR-M$_\star$ relation for eight cosmic surveys. The data have been adapted into the M$_\star$-sSFR plane in some cases. The colour of each line corresponds to the average redshift of its bin and the redshift colour scale is the same for all panels. The dotted lines correspond to M$_\star$ intervals where sample completeness is not guaranteed as reported in each work. The lower right corner of each field indicates the median scatter of the data.} \label{fig:surveys}
\end{figure*}

In Fig. \ref{fig:surveys} we show the evolution of the sSFR-M$_\star$ relation using eight different cosmic surveys:
\begin{enumerate}
    \item \cite{Perez-Gonzalez2008b} (PG08) selects galaxies from the \textit{Hubble} Deep Field-North \citep[HDF-N;][]{Williams1996}, Chandra Deep Field-South \citep[CDF-S;][]{Szokoly2004} and the Lockman Hole Field \citep[LHF;][]{Lockman1986} using \textit{Spitzer}-IRAC 3.6 $\mu$m and 4.5 $\mu$m bands. These observations are complemented by I-band ground-based observations to add galaxies which are faint in NIR. The redshift, M$_\star$ and SFR were calculated via spectral energy distribution (SED) fitting of photometric measurements in the ultraviolet (UV), optical and near- and mid-infrared (NIR/MIR).
    
    In their fig. 8, they show the redshift evolution of the sSFR for five M$_\star$ bins, which we transformed into the sSFR-M$_\star$ relation measured at six redshift values.
    
    \item \cite{Karim2011} (K11) selects galaxies in the Cosmic Evolution Survey \citep[COSMOS;][]{Scoville2007} fields from their \textit{Spitzer}-IRAC 3.6 $\mu$m emission. The redshift is taken from the catalogue in \cite{Ilbert2009}, determined using far-UV (FUV) to MIR photometry, the M$_\star$ was estimated with a parametric SED fit to the photometry data, and the SFR was measured from the 1.4 GHz luminosity of the sources, measured with the Very Large Array (VLA). Their fig. 5 shows the evolution of the sSFR-M$_\star$ relation, so no conversion was required, and we selected only half of the redshift bins to avoid clutter in the figure due to having too many lines.

    \item \cite{Schreiber2015} (Sc15) selects galaxies from GOODS, Ultra Deep Survey \citep[UDS;][]{Galametz2013} and COSMOS fields, using $H$-band images except for GOODS-N which uses $K_s$ band images. The redshift is calculated using photometry from NUV-to-8$\mu$m bands, the M$_\star$ is measured using parametric SED fitting and the SFR is calculated as the sum of the SFR obtained from UV and IR luminosity values. In their fig. 10, they show the SFR-M$_\star$ relation for different redshift values, which we converted to sSFR values by dividing them by the M$_\star$.
    
    \item \cite{Salmon2015} (Sa15) selects galaxies from the Cosmic Assembly Near-infrared Deep Extragalactic Legacy Survey\citep[CANDELS;][]{Grogin2011,Koekemoer2011} GOODS-S field in H-band images. The redshift is calculated with a SED fit to FUV-MIR photometric measurements from the HST, \textit{Spitzer} and ground-based observations, while the M$_\star$, dust attenuation, stellar ages and SFH shape are calculated with a Bayesian approach to SED fitting to the same data. The attenuation is used to correct the UV emission to calculate the SFR with the stellar ages and SFH used to further refine the values. We adapt the M$_\star$ and SFR values shown in their fig. 11.
    
    \item \cite{Tomczak2016} (T16) uses the sample from the FourStar Galaxy Evolution Survey (ZFOURGE) IR survey \citep{Straatman2016} which selects galaxies on the CDF-S, COSMOS, and UDS fields in the $K_s$ band. The redshift and M$_\star$ of the galaxies are obtained with separate SED fits to the five 0.3-8 $\mu$m photometry bands using different codes. The SFR is measured using archived \textit{Spitzer} and \textit{Herschel} measurements and calculated combining rest-frame UV and IR emission. In their fig. 3 they show the SFR-M$_\star$ relation.
    
    \item  \cite{Santini2017} (S17) uses data from the HST Frontier Fields \citep{Lotz2017} of four galaxy clusters, selecting the galaxies in the H-band image as well as adding faint detections performed on a weighted average of the $Y$, $J$, $JH$, and $H$  bands. The redshift is measured spectroscopically for a small portion of the sample and via SED fitting of optical-NIR photometry. The M$_\star$ was derived via parametric SED fitting and the SFR by correcting the UV emission for dust attenuation. In their fig. 2 they show the SFR-M$_\star$ relation.
    
    \item \cite{Barro2019} (B19) selects galaxies from the GOODS-N field in HST H-band images.
    The redshift is calculated via SED fitting using Survey of High-z Absorption Red and Dead Sources \citep[SHARDS;][]{Perez-Gonzalez2013} photometry and HST grism spectroscopy, which is roughly equivalent to low-resolution spectroscopy of R$\, \sim \,$50-210. The M$_\star$ is calculated via separate SED fits using UV-FIR photometry and the SFR is primarily based on rest-frame UV and IR emission which is corrected for dust-attenuation via information from several estimators depending on redshift. They give the values of SFR-M$_\star$ relation calculated with running medians in their table 15.
    
    \item \cite{Merida2023} (M23) combines the CANDELS data-set obtained in the five CANDELS fields, namely, the source catalogues from GOODS-N (B19), GOODS-S \citep{Guo2013}, COSMOS \citep{Nayyeri2017}, UDS, Extended Groth Strip \citep[EGS;][]{Stefanon2017}, with a sample of faint sources that they call SHARDS/CANDELS faint, performed in GOODS-N. They obtain the latter by combining observations from HST and SHARDS, looking for fainter sources via image stacking in the optical. Redshift, M$_\star$ and SFR values are computed similarly to B19. The sSFR-M$_\star$ relation we show in Fig. \ref{fig:surveys} uses the M$_\star$, SFR values from their table 6, which considers only the fraction of galaxies within their mass-complete limits.
\end{enumerate}

The main results in this article that we can compare with these studies are (i) the progressive steepening of the sSFR-M$_\star$ relation over time and (ii) the faster steepening above log M$_\star$/M$_\odot$ $\sim$10.5. The results related to morphology and present-day star formation cannot be easily reproduced using cosmic surveys.
As for (i), PG08, K11, Sc15 and T16 show this quite clearly, although only PG08, K11 and Sc15  clearly show (ii), the high M$_\star$ inflexion point at log M$_\star$/M$_\odot$ $\sim$10.5. PG08 shows the clearest change in slope at 10.5 in log M$_\star$/M$_\odot$, while K11 shows more of a peak in sSFR at this M$_\star$ for the earliest redshift, moving to lower M$_\star$ over time. Sc15 shows no clear transition point except for the two lowest redshift bins, where the relation becomes noticeably steeper at log M$_\star$/M$_\odot$ $\sim$10.5.

Regarding morphology as a key parameter for M$_\star$ growth, a companion publication to PG08 \citep{Perez-Gonzalez2008a} uses infrared data to measure the mass assembly rate of galaxies up to z$\sim$2 ($\sim$10 Gyr), and separates the sample into spheroidal and disc galaxies as proxies for early and late-type galaxies. They found a general decrease in sSFR over time with a difference between the two morphological groups, so that at the highest redshift values their sSFR is practically the same (within the uncertainty range), but they differ in more recent times, with the spheroidal group decreasing faster. There is a correlation between M$_\star$ and galaxy morphology, so this trend cannot be attributed to morphology alone. However, the general conclusion is that more massive earlier-type galaxies show a faster and stronger decline in sSFR compared to less massive later-type galaxies. This fits well with our results, although our sample and method allow us to break the correlation between M$_\star$ and morphology.

\section{Discussion}\label{sec:discussion}
The results shown here provide a general picture on how galaxies have grown their M$_\star$ over cosmic time and how this growth is regulated. Using the sSFR rather than the SFR makes it more easy to directly compare the SFH between galaxies as we are controlling for a general indicator of how 'big' a galaxy is: in statistical terms, we expect galaxies with higher M$_\star$ to have higher gas and halo masses, and also to be larger in physical size and have deeper potential wells.

It is argued that the SFMS is not a physical relation, but rather the result of other, more fundamental relations \citep{Lin2019,Sanchez2021a} such as the Schmidt-Kennicutt law \citep{Schmidt1959,Kennicutt1989} between the surface densities of molecular gas and SFR, and the molecular gas main sequence \citep[][]{Wong2013,Barrera-Ballesteros2020} between the surface densities of M$_\star$ and molecular gas.
Thus, to interpret the results, it is important to keep in mind that the correlations we detect do not necessarily imply a direct causal link between the parameters. Instead, they could arise from a combination of other, unobserved, correlations between more fundamental parameters in determining how galaxies grow. For example, many of the correlations with M$_\star$ are likely correlations with halo mass, but we still detect them due to the intrinsic correlation between halo and M$_\star$.
Alternatively, it is known that environment is a strong driver of galaxy evolution \citep[e.g.][]{Wetzel2012, Boselli2014} and strongly correlates with the morphology of the galaxies \citep[e.g.]{Martig2009, Schawinski2014, Smethurst2015}. It is quite feasible that environmental effects are at least part of the reason that, on average, earlier-type galaxies have a different evolution, as galaxies in more dense environments are both more likely to be early type and massive, and also have quenched their star formation earlier in their lifetime.

The accuracy of the SFH determination correlates with the age of the populations. The older a population is, the smaller the changes in its integrated spectrum due to the progressive loss of stars in a given interval. The most massive stars produce most of the luminosity, although they account for a much smaller fraction of the total M$_\star$ and they also die much faster than less massive stars. While the representative spectrum of a given population changes greatly between 0 and 500 Myr in age, the spectra at 10 Gyr and 10.5 Gyr in age, for example, are virtually indistinguishable.
The inescapable conclusion is that stellar population templates require wider sampling for older ages \citep[see ][]{Conroy2013,Walcher2011,Ibarra-Medel2016}. As a result, SFHs derived from stellar population synthesis become less reliable for the oldest ages and are generally unable to resolve the initial stages of galaxy formation. At the old end of the age range, the sampling of the SSP we employ has a gap of 1.7 Gyr between ages for the three oldest ones, which becomes a much higher 'error' in terms of redshift. The difference between 11 Gyr and 12.7 Gyr in LBT is 2.5 to 5 in redshift, highlighting one of the most striking differences between cosmic surveys and full spectral fitting regarding the measurement of when phenomena happen. Cosmic surveys lose sample completeness at higher redshift but in exchange gain ever increasing time resolution, while full spectral fitting always maintains complete sample integrity, but age measurements at high z are highly unreliable. This is compounded with the fact that older populations have much smaller light fractions, further adding to the uncertainty in age determinations. Despite this, full spectral fitting works quite well at finding differences in age between spectra assuming that the signal to noise, spectral resolution and spectral range are adequate. For this reason, while the absolute values of z at which galaxies quench are not robust, the observed differences between bins should be reliable.

For this reason, we consider the fact that several of the surveys show the same basic behaviour of the sSFR-M$_\star$ relation becoming flatter at higher LBT to be an agreement with our results, despite the fact that the quenching phase of the high-mass galaxies in the cosmic surveys occurs at significantly higher z than in our results. It is also worth pointing out that the surveys show significant differences between them in this regard, In PG08 the quenching occurs between z $\sim$2.5 and 1.75 whereas in K11 it occurs between z $\sim$1.8 and 1.1. Sc15 and T16 shows a steady steepening with no clear drop-off time. Our results suggest z$\sim$0.8, lower than either of the first two but not far from the lower value of 1.1 in K11. The inflection point at log M$_\star$/M$_\odot$ = 10.5 that we see coincides very well with that of PG08 and K11, but it is not clearly observed in any other survey. M23 is exempt as it does not reach beyond 10.5.
We also observe a general offset in the sSFR values between our results and that of cosmic surveys, likely related to some systematic offset in our measurements and at least partly explained by the fact that we do not have full coverage of the galaxies within the FoV of MaNGA (see Appendix \ref{sec:app-fraction}).

Despite this, even among the cosmic surveys there are significant differences in the sSFR. At z$\sim$0.5 and 1.2, for log M$_\star$/M$_\odot$ = 9.5, it varies between surveys up to 0.5 dex, while at higher z$\sim$4 it varies 1 dex or more between surveys. These differences should not be construed as these surveys being wrong, high-z measurements are difficult to constrain well and depend on the tracers of SFR employed. This is where our results can be useful despite the unreliability of the age determinations at high z and loss of resolution past z$\sim1-2$, as we can measure slight differences between sub-samples of galaxies consistently.

Compared to cosmological simulations, \cite{Behroozi2019} in their fig. 18 shows the evolution of the sSFR over cosmic time for different halo mass bins, where initially all mass bins have a similar sSFR value and evolve in parallel, but the more massive galaxies gradually drop in sSFR. This is consistent with our results regarding the fact that galaxies used to grow apace but more massive ones dropped off the SFMS. \cite{Dave2016} similarly shows in their fig. 6 the evolution of the sSFR-M$_\star$ relation in MUFASA simulations which goes from a negative relation with a bend at log M$_\star$/M$_\odot$ $\sim 10.5$ to a fairly flat (slightly positive correlation even) at high redshift. In the same figure they find the same behaviour comparing to EAGLE \citep[][]{Crain2015,Schaye2015} and ILLUSTRIS \citep{Genel2014,Vogelsberger2014} simulations. Like the comparison to the cosmic surveys, the redshift values at which the quenching occurs does not match our measurements well, which coincides much better with the results from cosmic surveys. This once again points towards the cosmic clock in our results being unreliable compared to cosmic surveys.

\subsection{The correlation at low M$_\star$} \label{sec:lowM_corr}
In the low end of the log M$_\star$ range we find a steeper correlation between M$_\star$ and sSFR below log M$_\star$/M$_\odot \sim 9$ for the intermediate LBT values 3-7 Gyr. This is an interesting feature if real, as it suggests that these very low M$_\star$ galaxies have a difference in the pace at which they grow their M$_\star$. It could be the result of either an increase in star formation efficiency for lower M$_\star$ galaxies or an increase in the ratio between molecular gas mass and stellar gas mass for lower M$_\star$ galaxies, that is, a break in the molecular gas main sequence (MGMS). For the latter case, it could once again arise from either a generally larger amount of gas or a higher efficiency in converting HI to H2. Lower M$_\star$ galaxies are expected to be heavily affected by stellar feedback expelling and heating the gas present in them so the explanation that these galaxies have higher molecular gas mass ratios fits better with the current paradigm on galaxy evolution models.

In Fig. \ref{fig:surveys} we showed the results from surveys and in several of them (Sa15, Sa17 and B19) this steepening of the relation at early times can be observed, especially for B19. It generally occurs at higher redshift values compared to our results (z$\sim0.2-0.8$), B19 shows it most clearly at z$\sim2-4$, though at z=0.25 there is a slight steepening between log M$_\star$/M$_\odot \sim 8-8.75$ with an increase of about 0.27 dex.
This feature, however, is prominent at the M$_\star$ ranges where the samples are not complete and it is typically explained away (and rightly so) as a selection effect arising from the fact that more star-forming galaxies tend to be brighter and as such in the fainter end of the relation they are preferentially detected. This selection effect grows for lower M$_\star$, fainter, and higher redshift galaxies and the negative correlation which becomes more prominent at early times arises as such.

For M23 the sSFR-M$_\star$ values we show correspond to the M$_\star$ range where completeness was guaranteed, but below the completeness limit many low to very low M$_\star$ galaxies are also reported. In their figure 13, the distribution in M$_\star$-SFR for the lowest redshift appears bimodal, with one population following the SFMS and another parallel but above it. The double peak in the distribution quickly fades at higher redshift values, leaving only the distribution above the SFMS. Including galaxies from their parent sample that are less massive than those included between their M$_\star$ limits would give similar results of a flattening of the SFMS or steepening of the sSFR-M$_\star$ relation. It bears mention, however, that this bimodality is only reported above z=1 and therefore it again does not exactly coincide with our results regarding when it is observed. Given that z=1 is the lowest z bin it is possible that it persists at later times, but it is not guaranteed.

Our results, by contrast, are free of the selection effects present in cosmic surveys as all galaxies are selected in the Local Universe. The MaNGA sample is reported to be mass-complete above $5\times10^8$ M$_\odot$ \citep[][]{Wake2017}. Even if the mass-completeness limit was underestimated and there were SFR selection effects in the MaNGA sample, our result at low M$_\star$ cannot be reproduced. This is because the selection effect only applies to the most recent sSFR value for each galaxy rather than at specific LBT values. Despite this, the spectral fitting procedure we employ is affected by degeneracies and method inaccuracies as described in the previous sub-section. It is therefore possible, especially as these galaxies have lower observed flux due to being in the lower M$_\star$ end, for this feature to be method-related, meaning we have two scenarios:
\begin{itemize}
    \item We observe a coincidental agreement between the effects of sample selection in cosmic surveys and method degeneracies from our spectral fitting.
    \item There is an unreported correlation for low-M$_\star$ galaxies which has so far been attributed entirely to selection effects intrinsic to cosmic surveys.
\end{itemize}

Further explorations are required to confirm whether this feature is real and understand the underlying reason for what would effectively be a flattening of the SFMS at very low M$_\star$. Observations with JWST and in the near future with instruments such as MOSAIC in the Extremely Large Telescope will provide constraints for the evolution of low M$_\star$ galaxies in cosmic time.

\subsection{Compatibility with theoretical models}
\subsubsection{The golden mass}
The golden mass is an explanation proposed by \cite{Dekel2006,Dekel2019} for the existence of a bi-modality in several parameters of galaxies mainly related to distribution functions and star formation, such as the one we observe in the SFMS and sSFR-M$_\star$ relation currently (see Fig. \ref{fig:sfms_current}). The basic idea is that star formation in galaxies is supported by the accretion of gas from the surroundings, but above $\sim 10^{12}$ M$_\odot$ in halo mass ($\sim 10^{10.5}$ M$_\odot$ in M$_\star$) the halo becomes virialised and the infalling gas is shock-heated as it approaches the galaxy, dramatically slowing down the accretion and conversely the SFR in galaxies.

This is supported by results from simulations \citep[][]{Birnboim2003,Keres2005,VandeVoort2012} and the bi-modality is explained as a transition in the dominant mechanism that regulates star formation inside the galaxies. Below log M$_\star$/M$_\odot \sim 10.5$, stellar winds mainly from supernovae (SNe) suppress star formation in an essentially self-regulating mechanism which becomes less efficient for higher masses as the growing gravitational potential makes it harder for the winds to expel gas from the galaxy \citep[e.g.][]{Silk1997,Hopkins2011,Hopkins2014}. Conversely, AGN feedback starts to become efficient at about the same M$_\star$ ($\sim 10^{10.5}$ M$_\odot$) \citep{Dekel2006} as, statistically, the central black hole mass grows with that of the host galaxy.
The outflow efficiency bimodality has been clearly detected and attributed to these two mechanisms (SNe and AGN) in simulations \citep{Mitchell2020}.

In our results we observe the bimodality clearly in the current properties, but we also see the signature of the golden mass in the fact that at precisely this mass value we see a bend in the relation for most LBT values. This bend appears even for currently star-forming galaxies and is observed even at the highest LBT we consider, albeit to a lesser degree than at the latest LBT such as 1 and 3 Gyr (see Fig. \ref{fig:ssfms_dist}).

The difference in the extent to which the bi-modality is observed in our results could be due to either a loss of resolution in our results for earlier times, which is expected, or a sign that the virialisation of the halo is a process which takes time to have an effect on the galaxies.
Overall, our results largely support a scenario where galaxies are initially self-regulated such that their sSFR is similar, regardless of their M$_\star$ until AGN feedback quenches galaxies above the golden mass.
The previously discussed feature of a negative correlation for low M$_\star$ galaxies is, however, at odds with the interpretation that below $\sim 10^{10.5}$ M$_\odot$ in M$_\star$, galaxies are dominated by stellar feedback. This feedback is expected to be especially efficient for very low M$_\star$ galaxies with shallow potential wells \citep[e.g.][]{Veilleux2005,Bland-Hawthorn2007,Mitchell2020} and therefore we would expect lower, not higher, sSFR for them. As mentioned above; to obtain a complete picture of this feature, we need to obtain measurements of the actual gas present in the galaxies as well as confirm the trend with deeper observations.

\subsubsection{Morphological quenching}
To collapse and form stars, the gas must cool and dissipate kinematic energy (especially turbulence). Consequently, any mechanism that adds turbulence to the ISM can suppress the ability of clouds to form stars and thus reduce the efficiency of star formation.
AGN and stellar feedback can directly introduce turbulence into the ISM via injection of kinetic energy, but galaxy dynamics can also create it through shear.

The most important of these mechanisms is the effect of the bulge on the disc, which can transfer a large amount of kinetic energy as turbulence and stabilise the disc to suppress star formation as it grows: the so-called morphological quenching \citep{Martig2009}. Bars can be considered as a kind of extension of the bulge into the disc and are expected to help amplify turbulence in the ISM around them. Apart from this, simulations show that the bars are able to funnel gas out of the disc into the centre of the galaxy, promoting star formation there, but ultimately also contributing to extinction by accelerating the time scale for the depletion of the gas \citep{Binney1991,Athanassoula1992,Boone2007,Sormani2015}.

Our results are very much in line with the idea that quenching and the way it is implemented in galaxies correlates with morphological type. We find differences in the number of galaxies that form stars, the extent to which the sSFR has decreased over cosmic time, and the timing of quenching.

\subsubsection{Galaxy interactions}
Mergers and interactions can have dramatic effects on the evolution of galaxies and will therefore be a key factor for it, especially in hierarchical galaxy formation models where today's galaxies are the result of many merger processes \citep[e.g.][]{White1978,White1991,Cole2000}.

Unfortunately, our results are largely blind to the effects of mergers because we measure the history of galaxies in terms of their current stellar composition. Any object in our sample that resulted from a previous merger of two galaxies has an SFH that is the sum of the two galaxies at any point in time. As a result, any such galaxy will appear in our figures at both a higher M$_\star$ and a higher SFR, keeping it within the global relation but extending it to higher M$_\star$. In general, we do not expect this to produce significant differences in our results, with one possible exception. Major mergers between galaxies tend to produce a galaxy that is more early type than its progenitors \citep[e.g.][]{Mihos1996,Lotz2008}. Additionally, given that a merged galaxy will be measured at higher M$_\star$ in the past than it should be (as two smaller galaxies), mergers are a possible explanation for the M$_\star$ segregation with morphology observed at very high LBT. The extent to which the segregation is due to mergers cannot be readily deduced from our data. \cite{Sanchez2019} measures the cosmic density of SFR at different redshifts using similar methodology to ours and finds that mergers do not significantly affect the results.

\section{Summary and conclusions}\label{sec:conclusions}
In this paper, we have studied how the current correlation between M$_\star$ and sSFR came to be by measuring both parameters at a different LBT using full spectral fitting. We have also studied whether and how the current star formation and morphology of the galaxies is tied to their past sSFR history. Lastly, we compared our results with several cosmic surveys to see if we can find an agreement with our findings.
Our main conclusions are:

\begin{itemize}
\item Galaxies in the past used to have similar mass-doubling times regardless of M$_\star$. Earlier-type and more massive galaxies then start to drop from the sequence around 7 Gyr ago in our results, producing the currently observed downsizing.
\item Galaxies that end up in different morphology bins in the present are segregated in the sSFR-M$_\star$ relation even at very early times. They are mainly separated in terms of their M$_\star$, but earlier-type galaxies also have a slightly higher sSFR high z; we measured a small ($\sim$0.1 dex) yet significant (median p-value for all M$_\star$ < $10^{-5}$) offset between early- and late-type galaxies.
\item The existence of a high-M$_\star$ inflection point is similar to predictions from simulations, though the z at which it occurs is much later in our results, likely due to methodology limitations.
\item The differences between morphology bins agree well with predictions for morphological quenching.
\item Comparing to cosmic surveys, we find agreement in several of them for some of the features we describe, though again the z values differ significantly.
\end{itemize}

These results show how SFHs derived through full spectral fitting can be used as a complementary study to cosmic surveys to probe cosmic star formation if its caveats are considered. The specific advantages of the method allow for a consistent measurement of the differences in the growth rate of distinct bins of galaxies. This allowed us to see how the morphology of the galaxies is determined by their past history starting from $>10$ Gyr ago.

\begin{acknowledgements}
We thank the referee for useful comments which improved the robustness of the results shown here.

ACF and PSB acknowledge financial support by the Spanish Ministry of Science and Innovation through the research grant PID2019-107427-GB-31.
RMM acknowledges support from Spanish Ministerio de Ciencia e Innovaci\'on MCIN/AEI/10.13039/501100011033 through grant PGC2018-093499-B-I00 and MDM-2017-0737 Unidad de Excelencia “Maria de Maeztu”-Centro de Astrobiología (CAB), INTA-CSIC, as well as by “ERDF A way of making Europe”. RMM acknowledges the support from the Instituto Nacional de T\'ecnica Aeroespacial SHARDS$^{JWST}$ project through the PRE-SHARDSJWST/2020 PhD fellowship, and by “ESF Investing in your future". SFS acknowledges funding from PAPIIT-DGAPA-AG100622 (UNAM) project.

SDSS is managed by the Astrophysical Research Consortium for the Participating Institutions of the SDSS Collaboration including the Brazilian Participation Group, the Carnegie Institution for Science, Carnegie Mellon University, Center for Astrophysics $\vert$ Harvard \& Smithsonian (CfA), the Chilean Participation Group, the French Participation Group, Instituto de Astrof\'{i}sica de Canarias, The Johns Hopkins University, Kavli Institute for the Physics and Mathematics of the Universe (IPMU) / University of Tokyo, the Korean Participation Group, Lawrence Berkeley National Laboratory, Leibniz Institut f\"{u}r Astrophysik Potsdam (AIP), Max-Planck-Institut f\"{u}r Astronomie (MPIA Heidelberg), Max-Planck-Institut f\"{u}r Astrophysik (MPA Garching), Max-Planck-Institut f\"{u}r Extraterrestrische Physik (MPE), National Astronomical Observatories of China, New Mexico State University, New York University, University of Notre Dame, Observat\'{o}rio Nacional / MCTI, The Ohio State University, Pennsylvania State University, Shanghai Astronomical Observatory, United Kingdom Participation Group, Universidad Nacional Aut\'{o}noma de M\'{e}xico, University of Arizona, University of Colorado Boulder, University of Oxford, University of Portsmouth, University of Utah, University of Virginia, University of Washington, University of Wisconsin, Vanderbilt University, and Yale University.

\end{acknowledgements}

%
%
\bibliographystyle{aa} 
\bibliography{export-bibtex} 

\begin{appendix} 
\section{Comparison between current SFR measured in H$\alpha$ and SSPs}

\begin{figure}
        \includegraphics[width=\columnwidth]{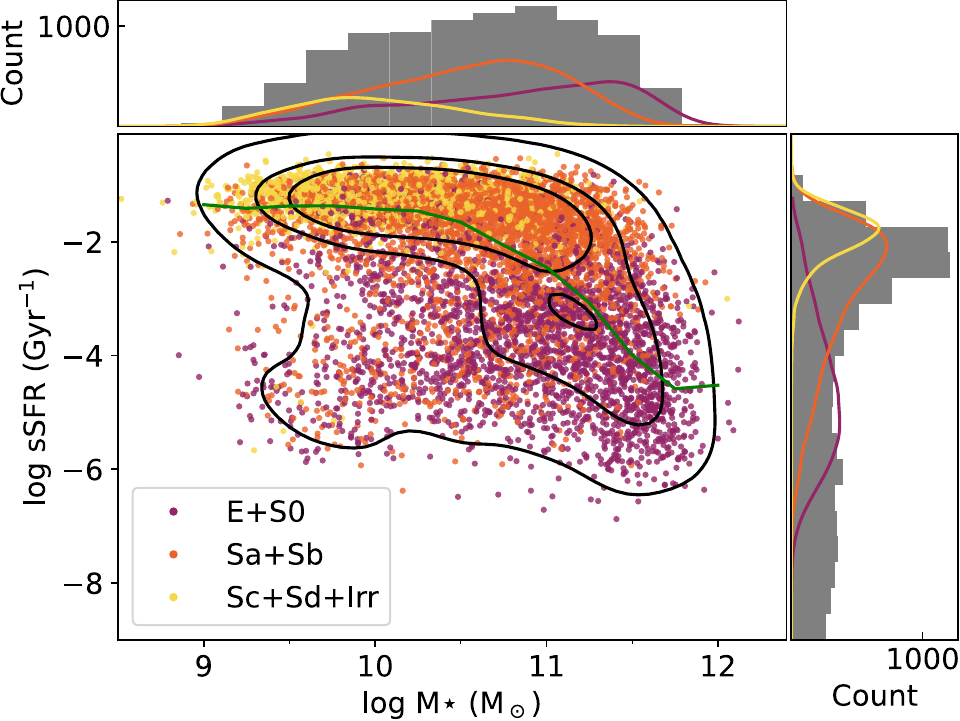}
    \caption{Currently observed sSFR of MaNGA galaxies compared to their M$_\star$ (bottom left), with the SFR measured using the H$\alpha$ emission line. The black contours enclose 95\%, 65\% and 35\% of the sample, respectively. The green line is the median sSFR within 0.25 dex wide M$_\star$ bins. In the top and right panels, the distributions in M$_\star$ and sSFR, respectively, are shown as histograms, with the three morphological bins shown in lines of the corresponding colour.}
    \label{fig:current_ha}
\end{figure}

As a base check for the reliability of the measurements using full spectral fitting and for comparison purposes we show here the current sSFR-M$\star$ relation using the SFR measured from the H$\alpha$ emission line, which can be found in Fig. \ref{fig:current_ha}.
The general shape of the relation is very similar to that of Fig. \ref{fig:sfms_current}, though the sSFR values reach significantly lower values for retired galaxies.

\begin{figure}
        \includegraphics[width=\columnwidth]{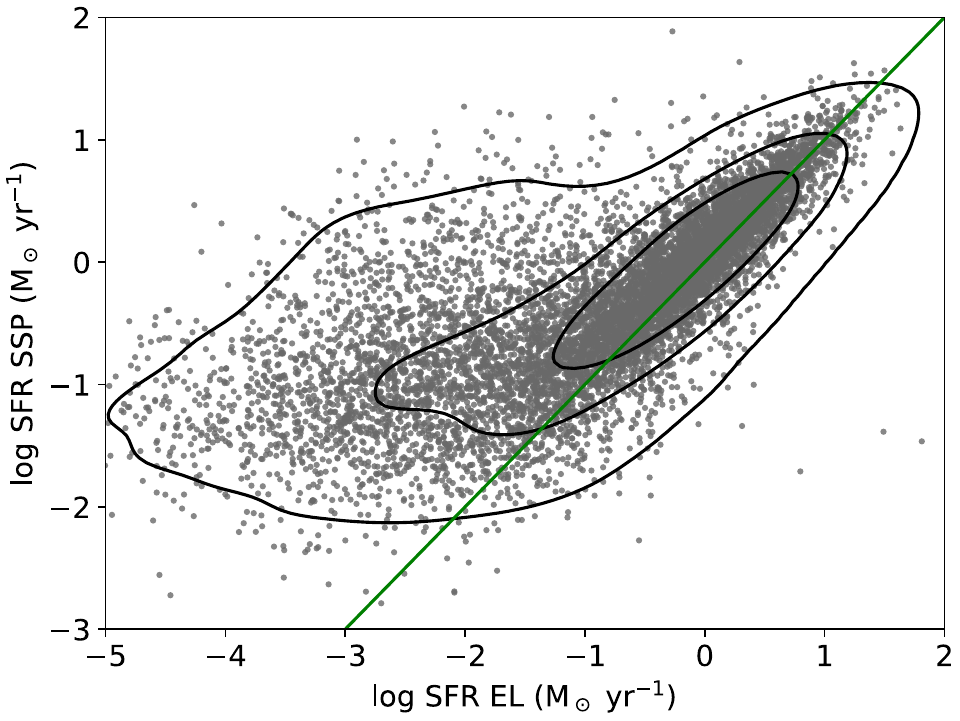}
    \caption{Comparison between the current SFR in our sample measured using the H$\alpha$ emission line (X-axis) and full spectral fitting selecting the last 30 Myr (Y axis). The black contours show the general distribution and the green line the 1:1 relation.}
    \label{fig:ha_vs_ssp}
\end{figure}

In Fig. \ref{fig:ha_vs_ssp} we show a direct comparison between the SFR measured using the H$\alpha$ emission line and the value from the stellar populations used in Fig. \ref{fig:sfms_current}. At relatively high values the two match very well, but for very low SFR the SSP values appear to have a floor. This is not a critical difference as it becomes relevant at around 0.01 M$_\odot$ yr$^{-1}$ which is a very low amount of star formation. The higher floor of the SSP values is likely due to method related reasons, such as the fact that, while the H$\alpha$ line can be measured in absorption, the minimum light fraction of a stellar population is 0, thus noise can only increase the SFR measurement when the actual value is 0.

\section{Impact of the FoV coverage in MaNGA}\label{sec:app-fraction}
Our measurements of the M$_\star$ and the SFR of the galaxies are intrinsically limited by the coverage that the MaNGA fibre bundles achieve for each galaxy. The MaNGA full sample is defined such that the primary sample, which makes up 2/3 of the galaxies, has $\sim80$\% of the galaxies covered out to 1.5 Re. The secondary sample, with 1/3 of the galaxies, conversely has 80\% of them covered out to 2.5 Re. Naturally, these are only selection targets and most of the galaxies will be covered to further galactocentric distances than these.

The resulting loss of light will impact our results, lowering the values of the SFR and M$_\star$ but the sSFR will be less impacted by this. Our results are mainly based on differences between bins of galaxies selected using M$_\star$, morphology and present-day star formation and, therefore, the main concern is that galaxies in different bins will have significantly different loss of light due to the FoV.

\begin{figure}
        \includegraphics[width=\columnwidth]{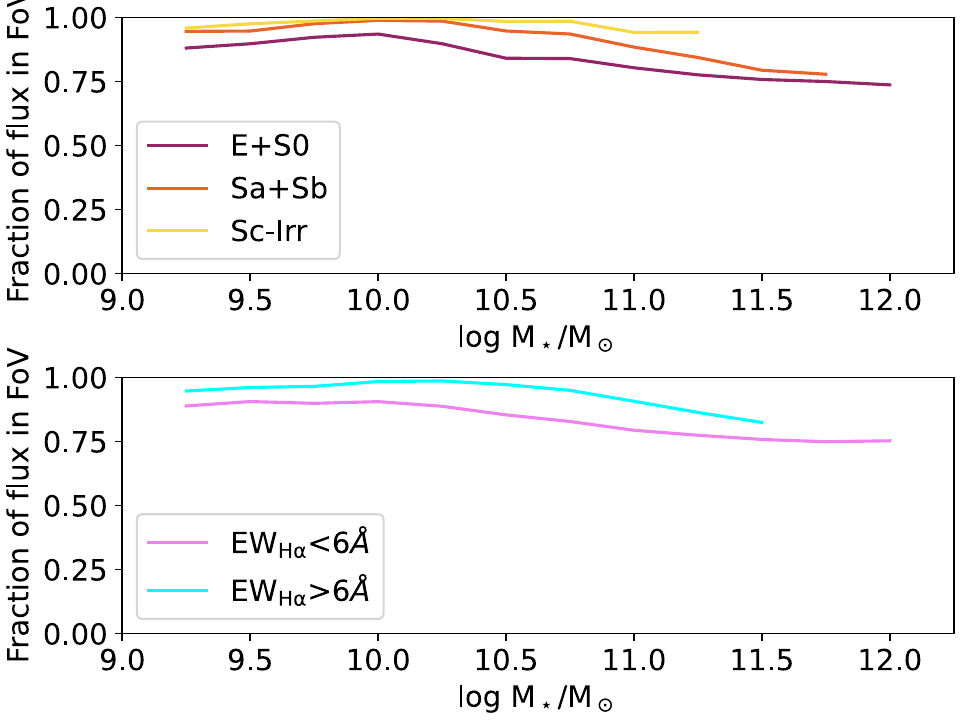}
    \caption{Average fraction of light contained within the FoV for each of the M$_\star$, morphology (top), and star formation (bottom) bins used in this article.}
    \label{fig:light_frac}
\end{figure}

To assess this, we have used the Sérsic index measurements from the NASA-Sloan Atlas Catalog and the FoV coverage measurements from \cite{Sanchez2022} to measure the average fraction of light that falls within the FoV by assuming the galaxies can be generally approximated using a Sérsic profile. The resulting average fraction for each of the bins considered is shown in Fig. \ref{fig:light_frac}. It bears mention that the M$_\star$ range does not reach values as low as those seen in Figs. \ref{fig:ssfms_dist} and \ref{fig:ssfms_morph}, due to the fact that the FoV is measured only in the present and, thus, the M$_\star$ range corresponds better to that of Figs. \ref{fig:sfms_current} or \ref{fig:current_ha}.

The figure shows that, in general, the average loss of light due to the FoV is small (0-20\%) and that it is fairly independent of M$_\star$, though there is a general trend to lower coverage in more massive galaxies.
There are, however, offsets between the morphology and star formation bins we use, such that earlier-type and retired galaxies have lower coverage.
Galaxies typically have age gradients such that the stellar populations are older in their centres and therefore the SFR values at high LBT will be less affected by the FoV if the FoV covers most of the galaxy. In any case, the higher loss of light for retired and earlier-type galaxies would mean a relative underestimation of the SFR at high LBT, but in Fig. \ref{fig:ssfms_dist} we show currently retired galaxies as being more star-forming, on average, than those that form stars in the present. Since the sSFR-M$_\star$ relation is quite flat the offset in M$_\star$ would also not significantly impact the results.

In Fig. \ref{fig:ssfms_morph} we show that earlier-type galaxies are located at slightly higher sSFR values and significantly higher M$_\star$ values comparatively. Similarly to Fig. \ref{fig:ssfms_dist}, the relative loss of flux in earlier-type galaxies means that these should be shifted to even higher M$_\star$ values and possibly to higher sSFR values.
As such, we conclude that the FoV coverage in MaNGA can have some effect on the measurements, but these effects cannot be the origin of the results we report.

\end{appendix}

%
%
\end{document}